\documentstyle[aps,manuscript]{revtex}

\begin{document}

\title{Re-visit the N/Z ratio of free nucleons from collisions of neutron
-rich nuclei as a probe of EoS of asymmetric nuclear matter}

\author {Qingfeng Li$^{1),}$
\footnote{Email: liqf@itp.ac.cn},
Zhuxia Li$^{2,1,3)}$ , Enguang Zhao$^{1)}$, H. St\"ocker$^{4)}$}
\address{
1) Institute of Theoretical Physics,
Chinese Academia of Sciences, P. O. Box 2735, Beijing 100080, P. R. China
\footnote{Mailing address}\\
2) China Institute of Atomic Energy, P. O. Box 275 (18),
Beijing 102413, P. R. China\\
3) Center of Theoretical Nuclear Physics, National Laboratory of
Lanzhou Heavy Ion Accelerator,
 Lanzhou 730000, P. R. China\\
4) Insitut f\"ur Theoretische Physik, Universit\"at Frankfurt, 60054, Germany\\}
\maketitle


\begin{abstract}
 The N/Z ratio of free nucleons from collisions of neutron-rich nuclei as a function of
 their momentum is studied by means of Isospin dependent Quantum Molecular Dynamics.
 We find that this ratio is not only sensitive to the form of the density dependence of the symmetry
 potential energy but also its strength determined by the symmetry energy coefficient.
 The uncertainties about the symmetry energy coefficient influence the accuracy of probing
 the density dependence of the symmetry energy by means of the N/Z ratio of free nucleons
 of neutron-rich nuclei.
 \\
\end{abstract}



Following the establishment of radioactive beam facilities at many
laboratories of different countries, the experimental studies on
the equation of state (EoS) for asymmetric nuclear matter become
possible. As is well known that the EoS for
asymmetric nuclear matter is one of the most important input for
astrophysics.

The EoS for asymmetric nuclear matter can be approximately expressed as
\begin{equation}
e(u,\delta )= \frac{\epsilon (u,\delta )}{\rho } = e_{0}
(u)+e_{sym} (u)\delta ^{2} , \label{eq1}
\end{equation}
where $u= \rho /\rho _{0} $ , $\delta = (\rho _{n} -\rho _{p}
)/\rho$ ; and $\epsilon (u,\delta )$ is the energy density;
$e_{0}$ is the energy per nucleon for symmetric nuclear matter and
$e_{sym}$ is the bulk symmetry energy. There exist large
uncertainties for the $e_{sym}$, especially, its density
dependence.
The symmetry energy at saturated normal density, i.e. the symmetry
energy coefficient, is not well constrained \cite{Pi02}.
The emperical value is $30\pm4$ MeV.
 The theoretically predicted value are rather different
from different approaches, its value is about 27-38 MeV by
non-relativistic Hartree-Fock approach \cite{Pe91}, 35-40 MeV by
relativistic mean field approach \cite{Ma81,Ru88,Fu95}, 31 MeV by
Brueckner-Hartree-Fock (BHF) theory \cite{Lee98}, 28.7 MeV by
extended  BHF theory \cite{Zuo99}, 26-34 MeV by relativistic BHF
\cite{Ha87,Mu87,Hu93}, 28.1 MeV by Brueckner-Bethe-Goldstone
approach \cite{My69}, {\it et al}. Furthermore, a recent study has shown
that the symmetry energy coefficient  increases as the isospin
asymmetry increases and the increasing  slope is quite different
for different versions of Skyrme force \cite{Br01}. At the same
time, we find rather different values are used in the
applications, for example, some authors \cite{Liu02,Li01} have
adopted values of 29 MeV and 31 MeV in the different IQMD model
calculations, whilst others \cite{Lib02,Chen02} used 30 MeV
and 35 MeV, respectively, in the framework of the IBUU model. 32
MeV was used in the stochastic Boltzmann-Nordheim-Vlasov (BNV)
calculations \cite{Ba02}, and 23.4 MeV was used in
\cite{Tsang01,Bo98}, {\it et al}.

The value of the symmetry energy
coefficient $a_{sym}$, which can be related to the strength of the symmetry
potential energy $C_S$ as

\begin{equation}
a_{sym} = \frac{C_{S} }{2} +\frac{\epsilon _{F}^{0} }{3} 
\label{eq2}
\end{equation}
if we
write the symmetry potential energy part in the form of
\begin{equation}
v_{sym} = \frac{C_{S} }{2} F(u) , \label{eq3}
\end{equation}
where F(u) gives the density dependence of symmetry potential
energy. One  can see from expressions 2) and 3) that the divergence of
the values of $a_{sym}$ means the uncertainties about the strength
of the symmetry potential energy. We have studied the influence of
the different $C_{S}$ on the isospin distribution of the emitted
nucleons, intermediate mass fragments, and light charged particles
and we have found this influence is obvious\cite{Li02}, while the
quantity concerned the most for the isospin dependent part of EoS
is its density dependence, which is one of the most important
input for the astrophysics. This has inspired people to find more
sensitive observable in order to pin down the form of the density dependence
of the EoS of the asymmetric nuclear matter. These works usually
adopted a fixed value of $C_{S}$ and then test the sensitivity of the
proposed sensitive observable to the different forms of the
density dependence of the symmetry potential energy term without checking
the influence of the strength of the symmetry potential. The aim
of this paper is to study how the existing uncertainties
concerning the $a_{sym}$ influence the sensitivity of the proposed
sensitive observable to the form of the density dependence of the
symmetry energy term. In this work, we mainly concentrate on the
momentum distribution of N/Z ratio of emitted nucleons in neutron
rich heavy ion collisions at energies ranging from several tens to
100A MeV since it has been proposed as a very sensitive observable
to the form of the density dependence of the EoS of asymmetric nuclear matter
\cite{Lib97}.

Concerning the form of density dependence of the symmetry potential energy, Prakash and Lattimer
has proposed \cite{La91,Pr88} as:
\begin{equation}
F(u)= \left\{\begin{array}{l}u  \\
u^{1/2}   \\
2u^{2} /(1+u)\end{array} \right.. \label{eq4}
\end{equation}
This form can be expressed as a simple one, i.e.,
\begin{equation}
F(u)= \left\{\begin{array}{l}u^{\gamma }   \\
2u^{\gamma } /(1+u^{\gamma -1} )
\end{array} \right..\label{eq5}
\end{equation}
In Fig. 1 we show F(u) as a function of reduced density with
$\gamma$= 0.5, 1.0, and 1.5, respectively. The $\gamma$= 0.5 case
corresponds to the asy-soft EoS and $\gamma$= 1.0 the asy-stiff
EoS, $\gamma$= 1.5 asy-super stiff EoS. From Fig. 1, one sees that
the type $F(u)= 2u^{\gamma } /(1+u^{\gamma -1} )$
is not too far away from linear density dependence in the densities given in the figure.
We then take a simple form, i.e., $F\left( u\right) = u^{\gamma }
$, in this work. It has been predicted that $\gamma$ is about 0.6,
based on a many-body theory calculations \cite{He00,Ak97}, which seems to be
 supported  by the recent experimental observation
\cite{Tsang01}.

\[
\fbox{Fig. 1}
\]

In this paper, we take C$_{S}$ to be 27, 35, and 50 MeV
(corresponding to the values of symmetry energy coefficient
a$_{sym}$ of 27, 31, and  38 MeV, respectively), which roughly
include the range of symmetry  energy coefficients predicted by
different theories and $\gamma$  to be 0.5, 1.0, and 1.5. 
The results with a reduced range of symmetry energy coefficients 
$C_S=37$, $40$, and $45$ (corresponding to $a_{sym}=32$, $34$, $36$) 
are also shown.
The
other parameters of EoS will be  shown in Table I. A soft EoS (K=
200 MeV) is used in the calculations.

\[
\fbox{Table I.}
\]

The compressibility contributed from symmetry potential can be
obtained by,

\begin{equation}
\begin{tabular}{l}
$K_{sym} \equiv 9\rho _{0}^{2} \frac{\partial ^{2} e_{sym} (u)}{\partial
^{2} \rho } |_{\rho = \rho _{0} }  $ \\
\ \ \ \ \ \ \ $= \frac{9}{2} \gamma (\gamma -1)C_{S} -\frac{2}{3}
\epsilon _{F}^{0}   .$
\end{tabular}
\label{eq6}
\end{equation}
Table II lists the K$_{sym}$ parameters when C$_{S}$= 27, 35, and
50 MeV, and $\gamma$= 0.5, 1.0, and 1.5. It shows that K$_{sym}$
changes  sign from negative to positive when $\gamma$ increases
from  0.5 to 1.5. At the same time, K$_{sym}$ increases for
$\gamma$= 0.5  and decreases for $\gamma$= 1.5 without changing
sign when  C$_{S}$ decreases from 50 MeV to 27 MeV. And these
values are well  within the wide range of K$_{sym}$ from about
-400 MeV to +466 MeV predicted by many-body theories
\cite{Lib98,Bo91}. Available data of giant  monopole resonances
does not give a stringent constraint on the  K$_{sym}$ parameter
either \cite{Sh93}.

\[
\fbox{Table II.}
\]

However, a complementary and perhaps more complete depiction of
the isospin dynamics can be obtained from the analysis of  the
density dependence of the neutron/proton chemical potential,

\begin{equation}
\begin{tabular}{l}
$\mu _{n/p} \equiv \frac{\partial \epsilon (u,\delta )}{\partial \rho
_{n/p} }  $ \\
\ \ \ \ \ \ $= \alpha u+\beta u^{\sigma } +\epsilon _{F}^{0}
u^{2/3}  $
\\
\ \ \ \ \ \ \ \ $+\lbrack \frac{C_{S} }{2} (\gamma -1)u^{\gamma }
-\frac{1}{9} \epsilon _{F}^{0} u^{2/3} ]\delta ^{2} \pm \lbrack C_{S}
u^{\gamma } +\frac{2}{3} \epsilon _{F}^{0} u^{2/3} ]\delta  .$
\end{tabular}
\label{eq7}
\end{equation}
From this equation, we find the difference between neutron and
proton chemical potential is
\begin{equation}
\mu _{n} -\mu _{p} = 4e_{sym} \rho \delta . \label{eq8}
\end{equation}
This expression indicates that the difference between $\mu_{n}$
and $\mu_{p}$ depends on both isospin asymmetry and the density.
When a more neutron-rich (or deficient) system is chosen, the
difference becomes more obvious. Since in the following we will
study reactions of $^{96}Zr+^{96}Zr$ and  $^{132}Sn+^{132}$Sn in Fig. 2 we
show $\mu_{n}$ and
$\mu_{p}$  versus density with C$_{S}$= 35 MeV, $\gamma$= 0.5,
1.5, and  C$_{S}$= 27 MeV, $\gamma$= 0.5 case for these two reactions, respectively.
The left panel is for  $^{96}$Zr+$^{96}$Zr  ($\delta$= 16/96),
the right one is for $^{132}$Sn+$^{132}$Sn ($\delta$= 32/132),
respectively. Fig. 2 shows that the larger the C$_{S}$ is, the
larger  the difference between $\mu_{n}$ and $\mu_{p}$ is and
accordingly  the more neutrons are emitted. Concerning the change of the form of
density dependence, when the
density is higher than the normal density, the difference between
$\mu_{n}$ and $\mu_{p}$  for asy-stiff  and asy-super stiff is larger than that for asy-soft
while, when density is lower than normal density, the tendency is just opposite.

\[
\fbox{Fig. 2}
\]

The single particle symmetry potential can be written as
\begin{equation}
\begin{tabular}{l}
$v_{sym}^{n/p} = \frac{\partial \left( v_{sym} (u,\delta) \rho \right)
}{\partial \rho
_{n/p} }  $ \\
\ \ \ \ \ \ $= \frac{C_{S} }{2} \lbrack \pm 2u^{\gamma } \delta
+(\gamma -1)u^{\gamma } \delta ^{2} ] ,$
\end{tabular}
\label{eq9}
\end{equation}
where $v_{sym} (u,\delta)$ represents the symmetry potential energy 
per nucleon.
One can see from Eq. 9 that the symmetry potential is repulsive
for neutrons and attractive for protons. So the effect of the
symmetry potential is to make more free neutrons and less free
protons. The first term in the bracket gives major contribution.
And for the $\delta^2$ term in the bracket, its sign depends on
whether $\gamma$ is less than 1 or not. It reduces the repulsive
force for neutrons and increases the attractive force for protons
for $\gamma <1 $ case and opposite for $\gamma > 1$ case and
consequently the $\delta^2$ term may reduce the effect of first
term slightly. Furthermore the tendency of $u^{\gamma}$ factor is
different for $\gamma <1$ and for $\gamma >1$ as density increases
from sub-normal to above-normal densities; at the sub-normal
density the repulsive force for neutrons is stronger when $\gamma
<1 $ than that when $\gamma >1 $ and for the above-normal density,
it is just opposite. And therefor one expects that the N/Z ratio
of emitted nucleons in collisions of neutron-rich nuclei with
$\gamma<1$  is over with the $\gamma>1$ case. As for the change of
the strength of $C_{S}$, it is simply that the larger the $C_{S}$
is the higher the N/Z ratio of emitted nucleons is. Our numerical
results clearly show all these tendencies.

From Eq. 9 we obtain the time evolution of $\vec{r}_{i}$ and
$\vec{p}_{i}$ contributed by symmetry potential term

\begin{equation}
\dot{\vec{r}} _{i= n/p}^{sym} = 0 , \label{eq10}
\end{equation}

\begin{equation}
\begin{tabular}{l}
$\dot{\vec{p}} _{i= n/p}^{sym} = -\frac{C_{S} }{2\rho _{0} } [2\pm
4(\gamma -1)u^{-1} \delta +(\gamma -1)(\gamma -2)u^{-2} \delta
^{2}
]u^{\gamma -1} \partial \rho _{n/p}  $ \\
\ \ \ \ \ \ \ \ \ \ \ \ \ \ \ \ \ \ \ \ $+[ -2+(\gamma -1)(\gamma
-2)u^{-2} \delta ^{2} ]u^{\gamma -1} \partial \rho _{p/n}  .$
\end{tabular}
\label{eq11}
\end{equation}

The isospin dependent quantum molecular dynamics model (IQMD) is used in
the calculations \cite{Li01} and the construction of clusters is in terms of the conventional
coalescence model \cite{Kr85}, in which particles with relative
momenta  smaller than P$_{0}$ and relative distances smaller than
R$_{0}$ are  considered to belong to one cluster. In this work
R$_{0}$ and P$_{0}$  are taken to be 3.5 fm and 300 MeV/c,
respectively, following \cite{Zhang99}. In  addition, only the
cluster with reasonable proton number Z and  neutron number N is
selected in order to get rid of nonphysical  clusters.

The secondary deexcitation of primary hot fragments is not taken
into account in the present calculations. This should not change
the general conclusion of this work.

First let us study the influence of the different forms of F(u)
on the N/Z ratio when $C_{S}$ value is taken as a fixed value,
i.e., C$_{S}$= 35 MeV. Fig. 3a) and 3b) show the ratio of emitted
neutron numbers and proton numbers versus their momentum for
reactions of $^{132}$Sn+$^{132}$Sn at  50$A$ MeV (Fig. 3a)) and
100$A$ MeV (Fig. 3b)) calculated with F(u) being to be the form of
$u^{1/2}$, $u$, and $u^{2}$.
 Different characters  denote calculation results obtained with different form of F(u)
and the  lines are shown just for guiding the eyes. The case for
reactions of $^{132}$Sn+$^{132}$Sn at E= 50A MeV has been
calculated in Ref. \cite{Lib97}. Here, for simplicity, the  impact
parameter b= 2 fm is chosen for central collisions. From  Fig. 3,
the obvious dependence of the N/Z ratio of free nucleons  on the
form of F(u) is seen, especially for the N/Z ratio of the
energetic neutrons from reactions of $^{132}$Sn+$^{132}$Sn at  E=
50A MeV case. The N/Z ratio of free nucleons with asy-soft case
($F(u)=u^{1/2}$) is much higher  than those with the asy-stiff
(F(u)=u) and asy-super stiff (F(u)=$u^{2}$) cases, which is in
agreement with our expectations. However, our results are not as
pronounced as that obtained in \cite{Lib97} though the general
trend is the same, which may due to the model dependence.

\[
\fbox{Fig. 3}
\]

Fig. 3b) shows that for 100A MeV case the dependence of N/Z ratio
of low momentum free nucleons on the form of F(u) is as strong as
that of energetic nucleons, while for 50A MeV case shown in
Fig. 3a) only the N/Z ratio of the energetic nucleons is sensitive
to the form of F(u). The comparison between Fig. 3b) and
Fig. 3a) means that there is an advantage for taking beam energy
at 100A MeV to extract the information of the form of F(u) because
the number of low momentum nucleons are much larger than those of
energetic nucleons and consequently the value of N/Z ratio can
measured more accurately.

Now let's study the dependence of the N/Z ratio of nucleons on the strength of the symmetry
potential energy, which closely related to the symmetry energy coefficient.
Fig. 4a) and 4b) show the N/Z ratio of emitted nucleons for reactions $^{132}$Sn+$^{132}$Sn
and $^{96}$Zr+$^{96}$Zr at E= 50$A$ MeV calculated with F(u)=u but
different C$_{S}$ values, namely, with $C_{S}$ = 27 MeV and 50 MeV. In Fig. 4a) we show
the results for central collisions of $^{132}$Sn+$^{132}$Sn, and
in Fig. 4b) we show the results for peripheral
collisions of $^{96}$Zr+$^{96}$Zr.
These results are similar to our previous work on the central collisions of
$^{96}$Zr+$^{96}$Zr at 400A MeV
\cite{Li02}, in which the sensitivity of the momentum distribution of N/Z ratio of free
nucleons to the strength of the symmetry potential energy is also shown but not as pronounced as
shown at lower bombarding energies studied in this work.
And furthermore, for case of central collisions of $^{132}$Sn+$^{132}$Sn at 50A MeV,
the dependence of N/Z ratio of energetic nucleons on $C_{S}$ is very pronounced,
 for example, the N/Z ratio of  nucleons with  momentum of about 300 MeV/c calculated with
$C_{S}$=50 MeV is about 1.5 times larger than that with $C_{S}$ =27 MeV. Relatively the N/Z ratio for
peripheral reactions of $^{96}$Zr+$^{96}$Zr at E= 50$A$ MeV is reduced but the ratio of
(N/Z)$_{Cs=50}$/(N/Z)$_{Cs=27}$ is similar with that for central reactions of $^{132}$Sn+$^{132}$Sn at 50A MeV.
Fig. 4a) and Fig. 4b) show that the influence of strength of symmetry potential energy on th N/Z ratio
of free nucleons is as obvious as that of the form of F(u) shown in Fig. 3a) and Fig. 3b).

\[
\fbox{Fig. 4}
\]

In order to investigate the most pronounced and the least
pronounced sensitivity of N/Z ratio to the density dependence of
the symmetry potential energy we study the influences of the
different combination of both the strength and the form of the
density dependence of symmetry potential. We define ratios

$$R_{most}=\frac{N/Z(F(u)=u^{1/2},Cs=50)}{N/Z(F(u)=u^{3/2},Cs=27)}$$

and

$$R_{least}=\frac{N/Z(F(u)=u^{3/2},Cs=50)}{N/Z(F(u)=u^{1/2},Cs=27)}$$

Fig. 5 shows the calculation results of the N/Z ratio of emitted nucleons versus their momentum for
central collisions of $^{96}$Zr+$^{96}$Zr at 50A MeV with
F(u)=u$^{1/2}$, $C_{S}$=50 and 27 MeV and with F(u)=u$^{3/2}$, $C_{S}$=27 and 50
MeV, respectively. The $R_{most}$ and $R_{least}$ are shown in the right-top plot.
One can see that the $R_{most}$ is more pronounced, while the $R_{least}$ is much less pronounced (about
unit), than those
shown in Fig. 3 and Fig. 4 as well.
 Fig. 5 tells us that under the influence of the uncertainties of
 symmetry energy coefficient what sensitivity of the N/Z ratio of emitted nucleons
 in intermediate energy neutron-rich heavy ion collisions to the form of the density
 dependence of symmetry potential energy can be obtained.

\[
\fbox{Fig. 5}
\]

Recently, a much smaller range of $32$ MeV $\leq a_{sym} \leq 36$ MeV for symmetry energy at 
saturation (volume asymmetry) is deduced from the isovector GDR in $^{208}Pb$ and the available 
data of diffrences between neutron and proton radii in Ref. \cite{Vre03}. It is worthwhile to see
how the ratios of $R_{most}$ and $R_{least}$ look like with $32$ MeV $\leq a_{sym} \leq 36$ MeV. 
Fig. 6 shows the calculation results with 
F(u)=u$^{1/2}$, $C_{S}$=45 and 37 MeV and with F(u)=u$^{3/2}$, $C_{S}$=37 and 45
MeV, respectively. The corresponding $R_{most}$ and $R_{least}$ with $C_{S}$=37 and 45 MeV are also shown in the right-top plot. 
From this figure, one can find that even with this small range the sensitivities of the N/Z ratio of free nucleons to the density dependence of the symmetry energy is still influenced by the uncerntainties of $a_{sym}$ though the ratio $R_{least}$ is now larger than one which means the N/Z ratio of free nucleons is enhanced for $\gamma$=0.5 case. 
Our study shows that it is urgently needed to have a more precise value of the $a_{sym}$ in order to get more definite information of the density dependence of the symmetry energy.

\[
\fbox{Fig. 6}
\]

To summarize, in this paper we have studied the sensitivity of
the N/Z ratio of free nucleons in collisions of neutron-rich
nuclei at energies of 50A MeV and 100A MeV to the form of the
density dependence of the symmetry potential energy term and the
strength of the symmetry potential by using IQMD transport model.
We have found that the N/Z ratio of free nucleons are sensitive to
both the form of the density dependence of the symmetry potential
and the strength of the symmetry potential term as well. The
results of the influences of the different combinations of both
symmetry potential strength and the form of the density dependence
of symmetry potential show that the uncertainties of the symmetry
energy coefficient largely reduce the sensitivity of the N/Z ratio
of free nucleons from collisions of neutron-rich nuclei as a probe
of the form of the density dependence of the symmetric energy
part. It is urgently needed to have a more precise value of the 
$a_{sym}$ in order to get more definite information of the density dependence of the symmetry energy.

The work is
supported by the National Natural Science Foundation of China
under Nos. 10175093 and 10235030, Science Foundation of Chinese
Nuclear Industry and Major State Basic Research Development
Program under Contract No. G20000774, the Knowledge Innovation
Project of the Chinese Academy of Sciences under Grant No.
KJCX2-SW-N02, and the CASK.C. Wong Post-doctors Research Award Fund. 
Z. Li acknowledges the warm hospitality of the
Insitut f\"ur Theoretische Physik, Universit\"at Frankfurt.


\newpage

\begin{table}
\caption{ Parameters used in calculations.}

\begin{tabular}{|c|c|c|c|c|c|c|}
$\alpha (MeV)$ & $\beta (MeV)$ & $\gamma $ & $\rho _0 (fm^{-3})$ &
$K = (MeV)$ & $L (fm)$ & $C_{Yuk} (MeV)$ \\ \hline $-356$ & $303$
& $7/6$ & $0.168$ & $200$ & $1.45$ & $-5.5$ \\
\end{tabular}
\end{table}

\begin{table}
\caption{The $K_{sym}$ of different $C_S$ and $\gamma$.}

\begin{tabular}{|l|l|c|c|c|}
& $\gamma \ \ \backslash \ \ C_{S}^{{}}\ (MeV)$ & 27 & 35 & 50 \\
\hline & 0.5 & -55.7 & -64.7 & -81.6 \\ \cline{2-5} $K_{sym}\
(MeV)$ & 1.0 & -25.3 & -25.3 & -25.3 \\ \cline{2-5} & 1.5 & 65.8 &
92.8 & 143.4 \\
\end{tabular}

\end{table}

\newpage

\begin{figure}
 \caption{The dependence of F(u) on $\gamma$. The $\gamma$ value is
chosen 0.5, 1.0, and 1.5, respectively.}
 \end{figure}

\begin{figure}
 \caption{$\mu_{n}$ and $\mu_{p}$ as a function of u for
$\delta$= 16/96 and 32/132 with C$_{S}$= 35 MeV, $\gamma$= 0.5,
1.5, and C$_{S}$= 27 MeV, $\gamma$= 0.5 cases, respectively.}
 \end{figure}

\begin{figure}
 \caption{The density dependence of N/Z ratio of emitted nucleons
versus momentum. The reactions $^{132}$Sn+$^{132}$Sn at 50$A$ MeV
and 100$A$ MeV are chosen in a) and b), respectively. The different
line types are drawn only for guiding the eyes, as well as those
in the next figures. The ratio of N/Z with $\gamma=0.5$ and $1.5$
is shown in their right-top plots, respectively.}
 \end{figure}

\begin{figure}
 \caption{The dependence of N/Z ratio of emitted nucleons on symmetry
potential strength as a function of momentum. Hereinto a) chooses
 $^{132}$Sn+$^{132}$Sn reaction in central collisions, b) chooses
$^{96}$Zr+$^{96}$Zr reaction  in peripheral collisions,
respectively. The ratio of N/Z with $C_{S}=50$ MeV and $27$ MeV
is shown in their right-top plots, respectively.}
 \end{figure}

\begin{figure}
 \caption{The N/Z of nucleons versus momentum in central collision
$^{96}$Zr+$^{96}$Zr at 50$A$ MeV with different symmetry
potential: C$_{S}$= 27 and 50 MeV when $\gamma$= 0.5 and 1.5. 
The $R_{most}$ and $R_{least}$ (see text)
are shown in the right-top plot.}
 \end{figure}

\begin{figure}
\caption{The same as in Fig. 5 with different symmetry
potential. Here C$_{S}$= 37 and 45 MeV are chosen.}
\end{figure}

\begin{thebibliography}{99}
\bibitem{Pi02}J. Piekarewicz, Phys. Rev. {\bf C66}, (2002) 034305.\\
\bibitem{Pe91}J.M. Pearson, Y. Aboussir, A.K. Putt, Nucl. Phys. {\bf
A528}, (1991) 1.\\
\bibitem{Ma81}T. Matsui, Nucl. Phys. {\bf A370}, (1981) 365.\\
\bibitem{Ru88}M. Rufa, P.G. Reinhard, J.A. Maruhn, W. Greiner, M.R.
Strayer, Phys. Rev. {\bf C38}, (1988) 390.\\
\bibitem{Fu95}R.J. Furnstahl, H.B. Tang, B.D. Serot, Phys. Rev. {\bf
C52}, (1995) 1368.\\
\bibitem{Lee98}C.H. Lee, T.T.S. Kuo, G.Q. Li, G.E. Brown, Phys. Rev.
{\bf C57}, (1998) 3488.\\
\bibitem{Zuo99}W. Zuo, I. Bombaci and U. Lombardo, Phys. Rev. {\bf C60},
(1999) 024605.\\
\bibitem{Ha87}B. terHaar and R. Malfliet, Phys. Rev. Lett., {\bf 59},
(1987) 1652.\\
\bibitem{Mu87}H. M\"uller, M. Prakash and T.L. Ainsworth, Phys. Lett.
{\bf B199}, (1987) 469.\\
\bibitem{Hu93}H. Huber, F. Weber and M.K. Weigel, Phys. Lett. {\bf
B317}, (1993) 487.\\
\bibitem{My69}W.D. Myers and W.J. Swiatecki, Ann. Phys. (N.Y.), {\bf
55}, (1969) 186.\\
\bibitem{Br01}F.L. Braghin, Nucl. Phys. {\bf A696}, (2001) 413.\\
\bibitem{Liu02}J.Y. Liu, W.J. Guo, Y.Z. Xing, {\it et al}., Phys. Lett. {\bf
B540}, (2002) 213.\\
\bibitem{Li01}Qingfeng Li, Zhuxia Li, Phys. Rev. {\bf C64}, 
(2001) 064612.\\
\bibitem{Lib02}Bao-An. Li, Andrew T. Sustich, Matt Tilley, Bin Zhang,
Nucl. Phys. {\bf A699}, (2002) 493.\\
\bibitem{Chen02}Lie-Wen Chen, Vincenzo Greco, Che Ming Ko, Bao-An Li,
Phys. Rev. Lett. {\bf 90}, (2003) 162701.\\
\bibitem{Ba02}V. Baran, M. Colonna, M.Di Toro, V. Greco,
M.Zielinska-Pfabe and H.H. Wolter, Nucl. Phys. {\bf A703}, (2002) 603.\\
\bibitem{Tsang01}M.B. Tsang, W.A. Friedman, C.K. Gelbke, W.G. Lynch, G.
Verde, and H. Xu, Phys. Rev. Lett. {\bf 86}, (2001) 5023.\\
\bibitem{Bo98} A. Bohr and B.R. Mottleson, "Nuclear Structure, Vol II",
W.A. Benjamin Inc., (1998).\\
\bibitem{Li02}Qingfeng Li and Zhuxia Li, Mod. Phys. Lett. {\bf A17},
(2002) 375.\\
\bibitem{Lib97}B.A. Li, C.M. Ko, and Z. Ren, Phys. Rev. Lett. {\bf
78}, (1997) 1644.\\
\bibitem{La91}J.M. Lattimer {\it et al}., Phys. Rev. Lett. {\bf 66}, 
(1991) 2701.\\
\bibitem{Pr88}M. Prakash {\it et al}., Phys. Rev. Lett. {\bf 61}, 
(1988) 2518.\\
\bibitem{He00}H. Heiselberg and M. Hjorth-Jensen, Phys. Rep. {\bf 328},
(2000) 237.\\
\bibitem{Ak97}A. Akmal and V.R. Pandharipande, Phys. Rev. {\bf C56},
(1997) 2261.\\
\bibitem{Lib98}Bao-An Li {\it et al}., Int. Jou. Mod. Phys. {\bf E7}, 
(1998) 147.\\
\bibitem{Bo91}I. Bombaci and U. Lombardo, Phys. Rev. {\bf C44}, 
(1991) 1892.\\
\bibitem{Sh93}S. Shlomo and D.H. Youngblood, Phys. Rev. {\bf C47}, 
(1993) 529.\\
\bibitem{Kr85}H. Kruse {\it et al}., Phys. Rev. {\bf C31}, (1985) 1770.\\
\bibitem{Zhang99}F.S. Zhang {\it et al}., Phys. Rev. {\bf C60}, 
(1999) 064604.\\
\bibitem{Vre03} D. Vretenar, T. Nik\v si\' c, and P. Ring, nucl-th/0302070.\\
\end{thebibliography}
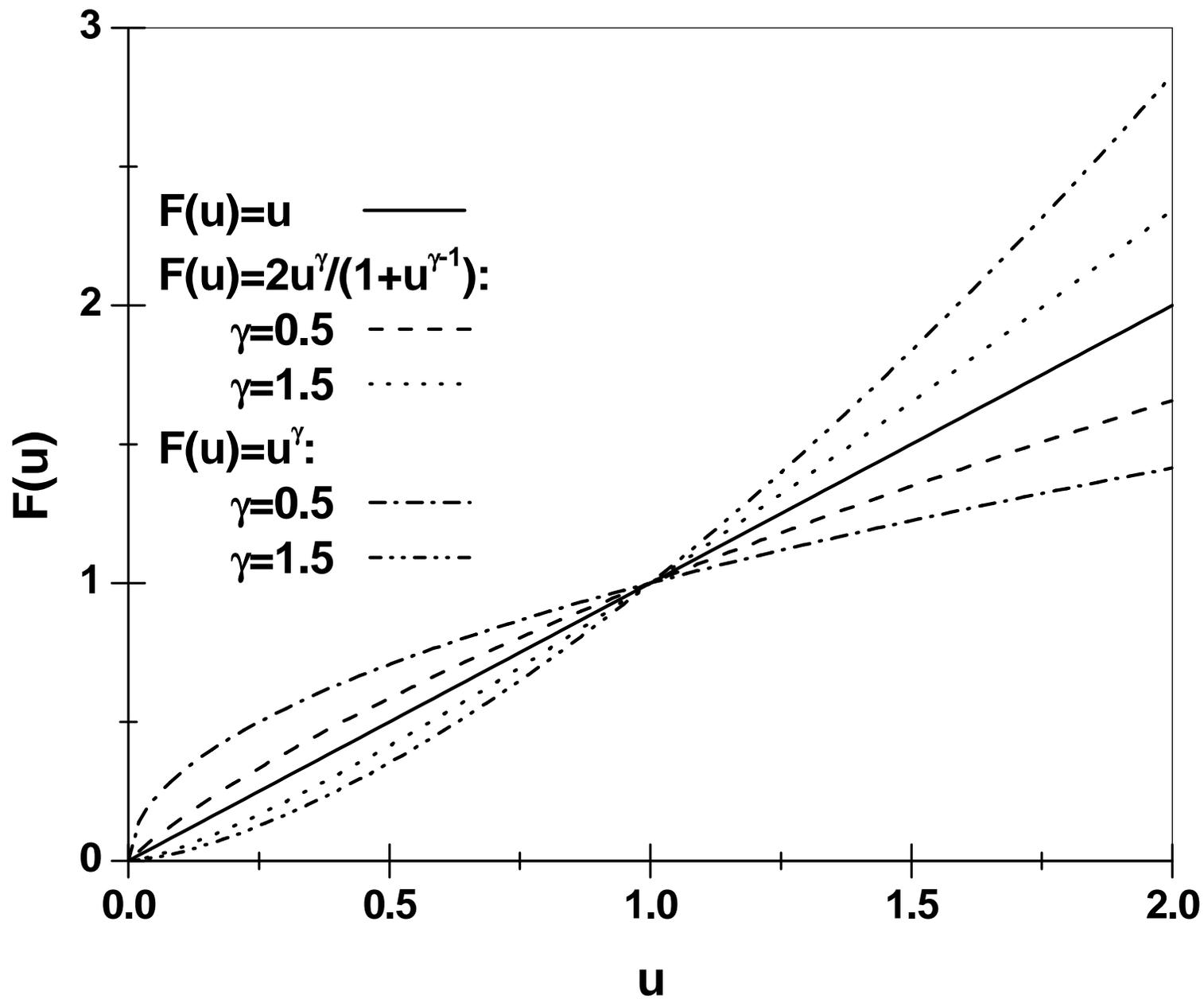
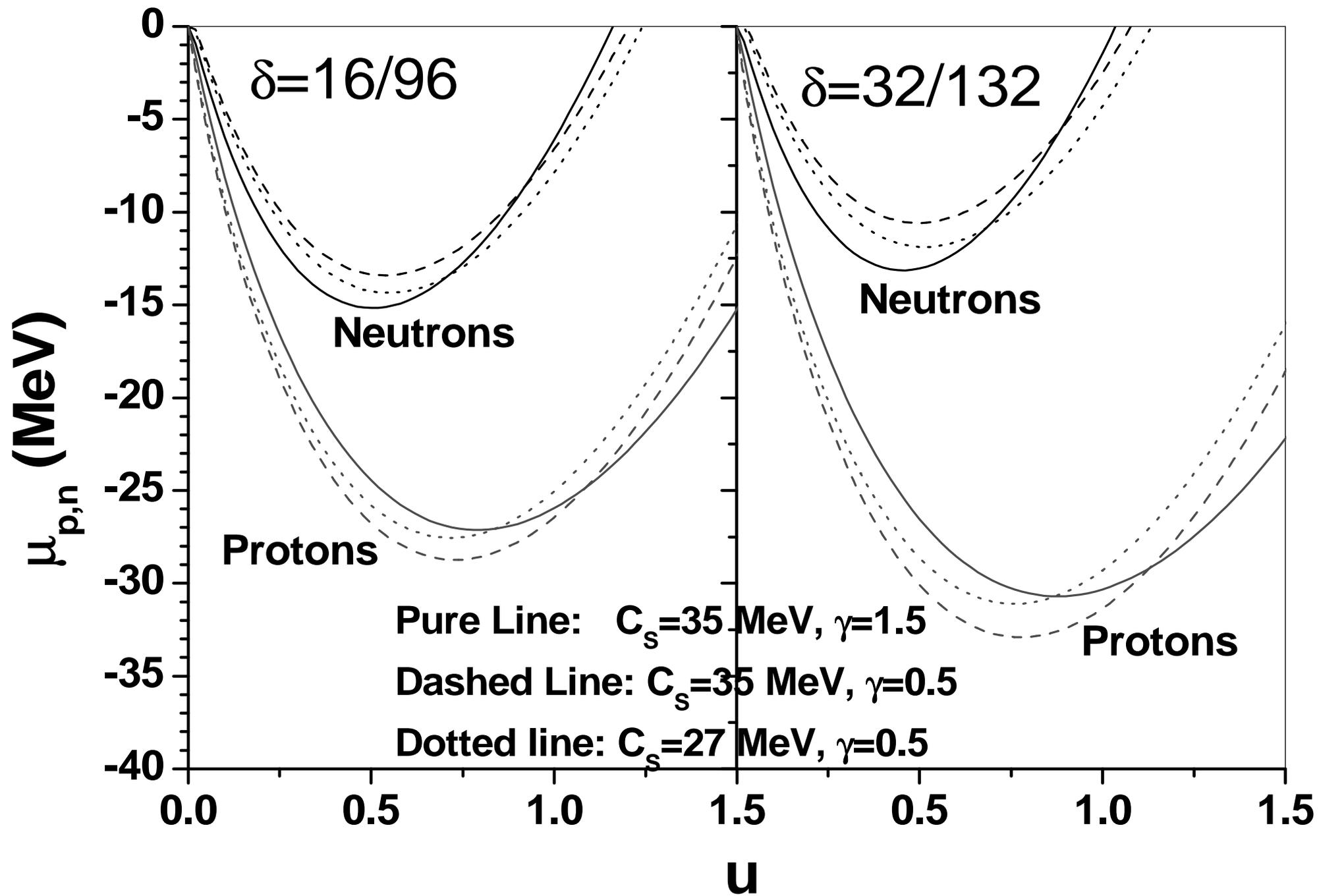
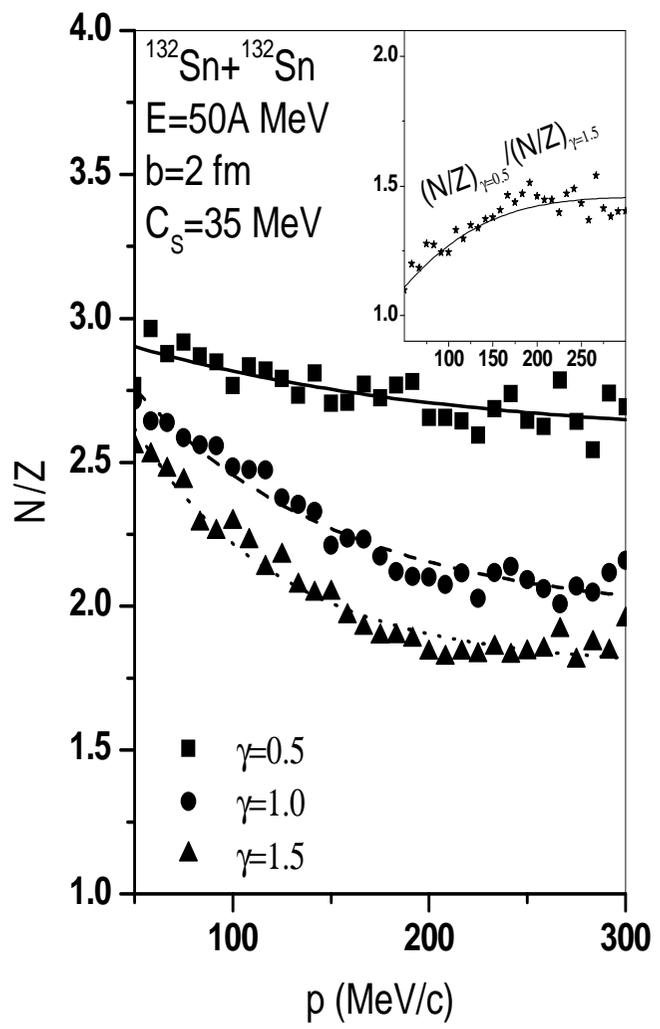
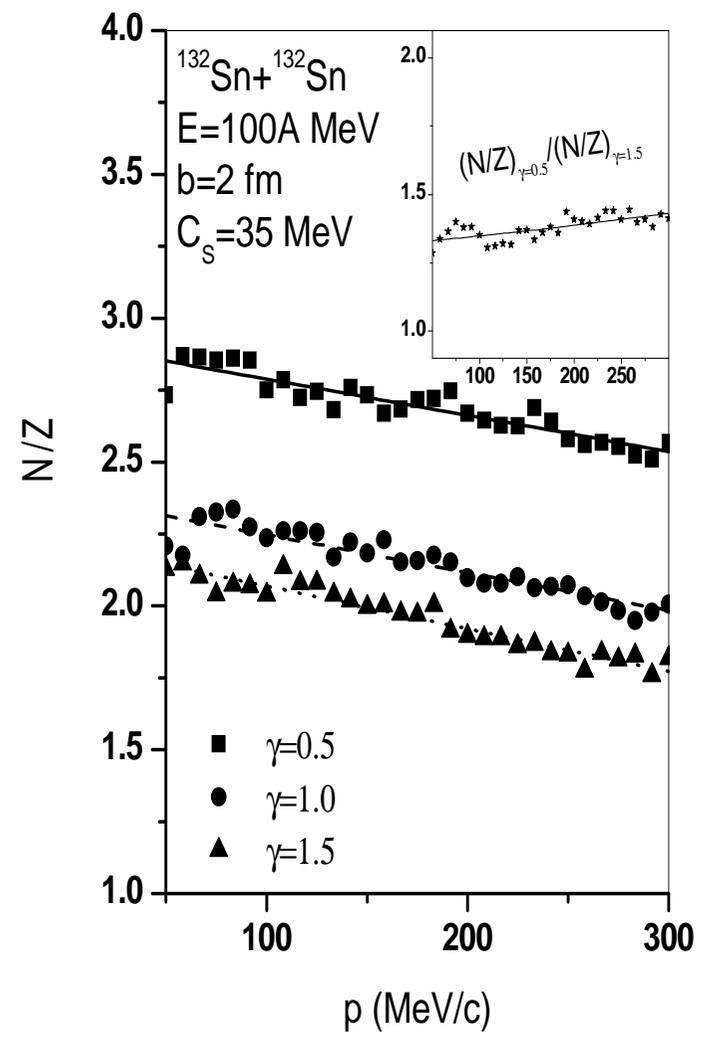
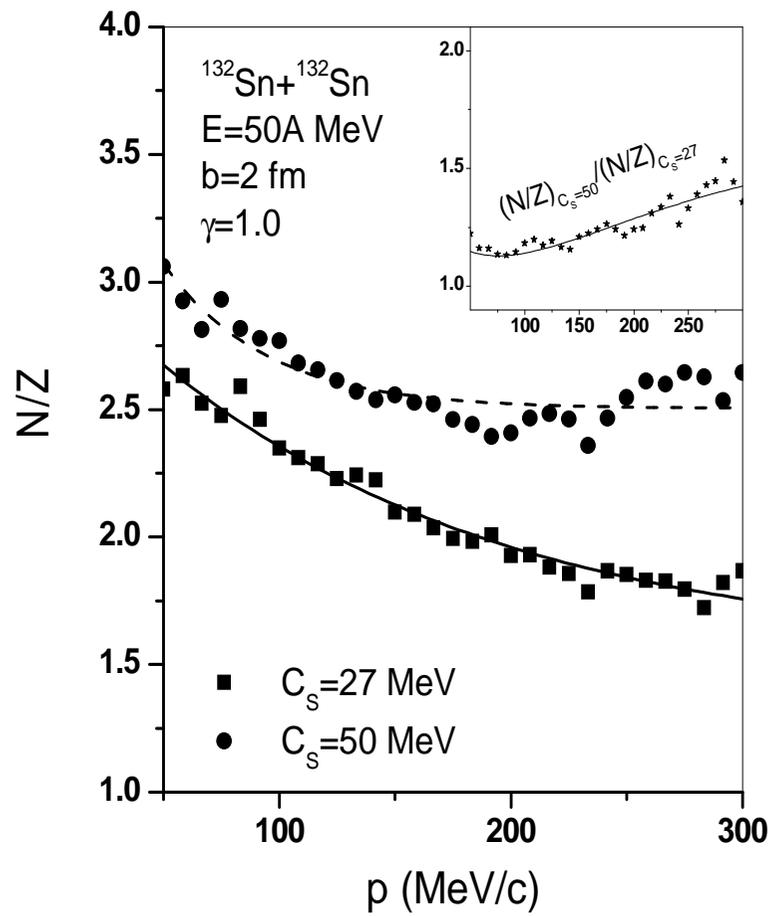
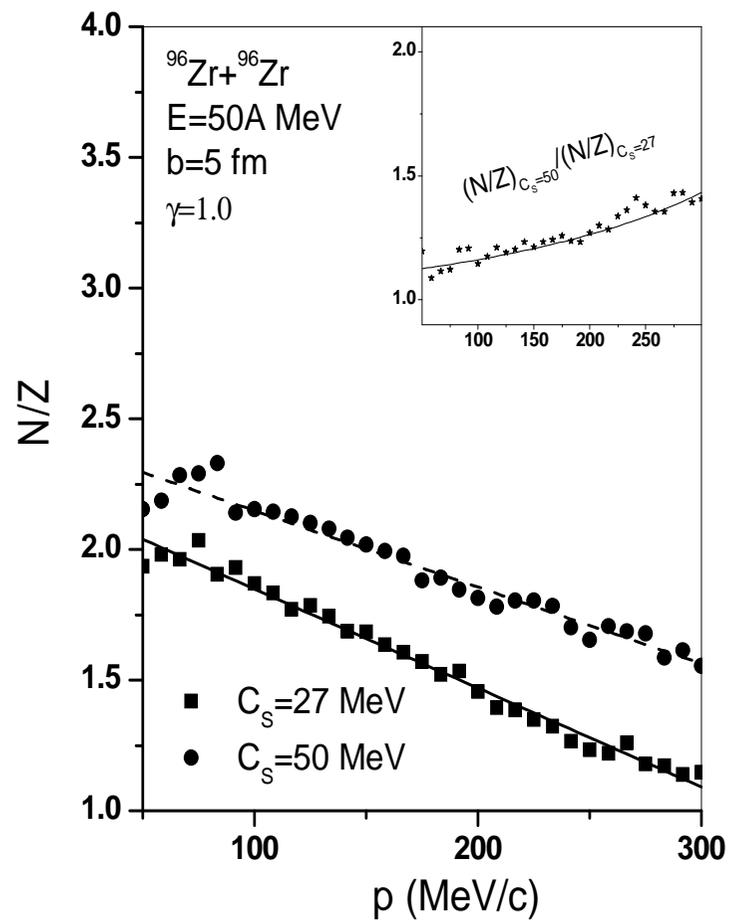
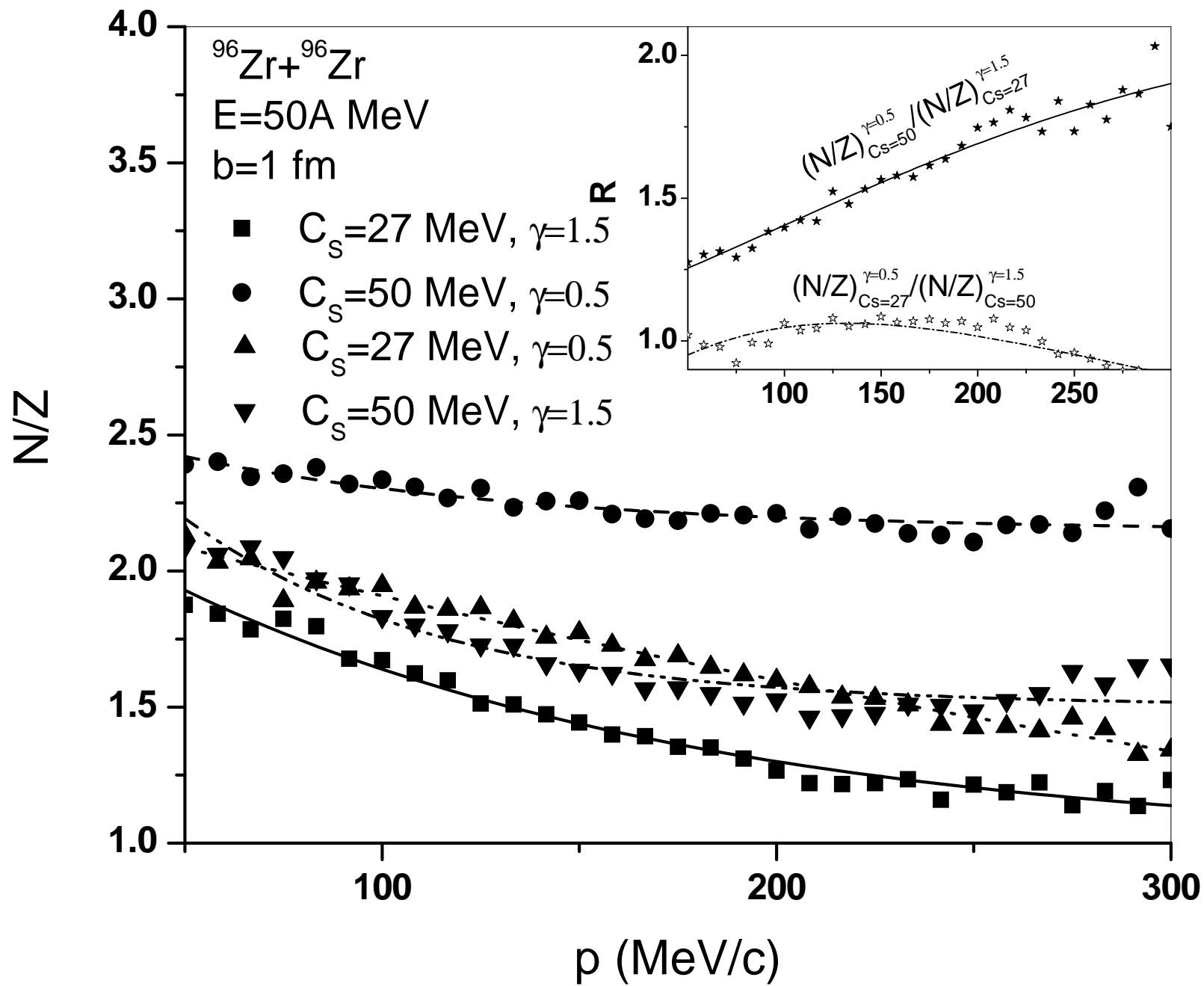
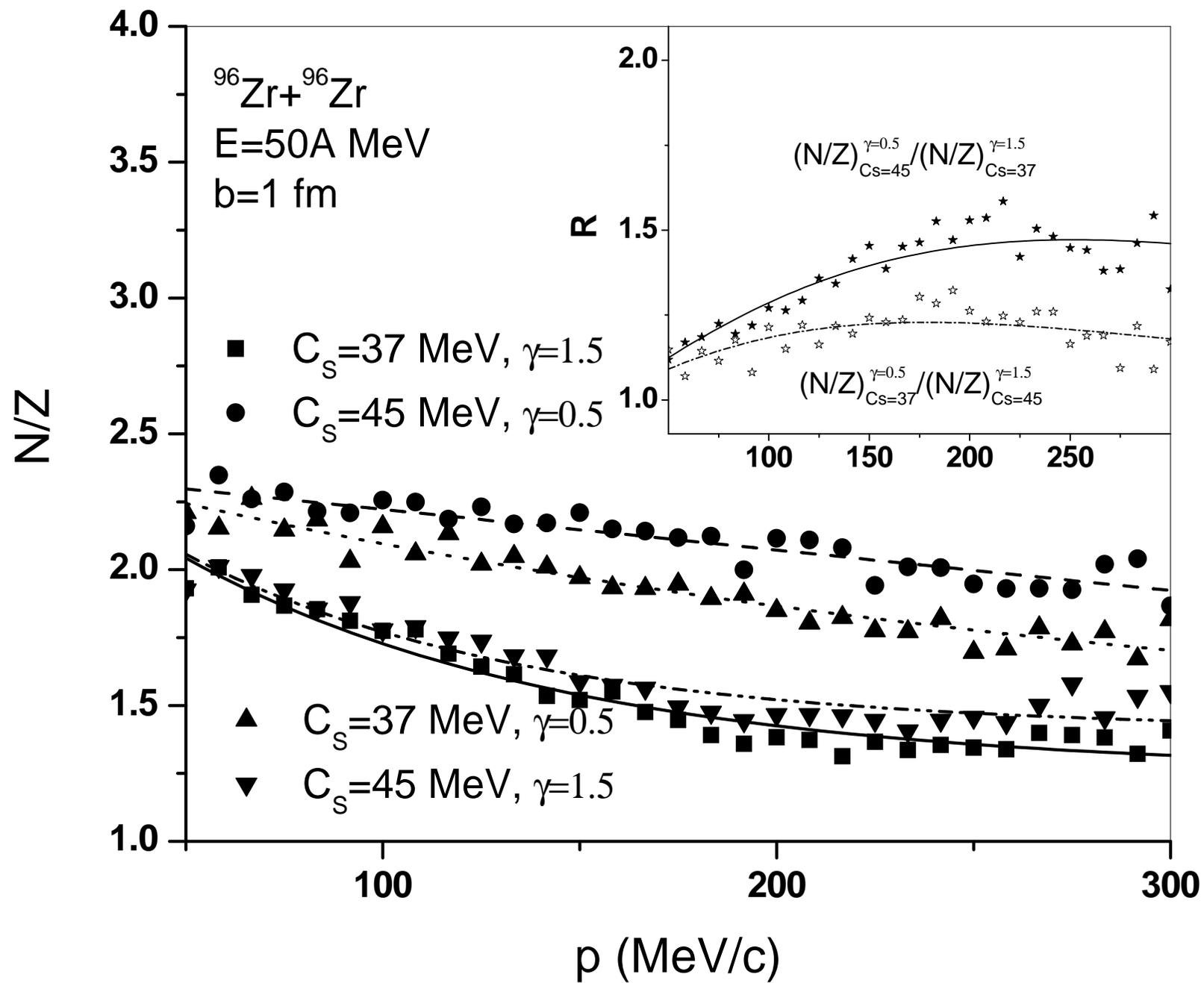
\end{document}